\documentclass[letters,fleqn,useAMS,usenatbib]{mnras}
\usepackage{graphicx}
\usepackage[english]{babel}
\usepackage{amsmath}
\usepackage{amssymb}
\usepackage{natbib}
\usepackage{wasysym}
\usepackage{newtxtext,newtxmath}
\usepackage[T1]{fontenc} 

\title[The CGM at 1~kpc uniform spatial resolution]{Cosmological simulations of the circumgalactic medium with 1~kpc resolution: enhanced H\,{\sc i} column densities}
\author[F.~van~de~Voort et al.]{Freeke van~de~Voort,$^{1,2}$\thanks{E-mail: freeke.vandevoort@h-its.org} 
Volker Springel,$^{3,1}$ Nir Mandelker,$^{2,1}$ 
\newauthor
Frank~C.~van den Bosch$^{2}$ and R\"udiger Pakmor$^{3,1}$ \\
$^{1}$Heidelberg Institute for Theoretical Studies, Schloss-Wolfsbrunnenweg 35, 69118, Heidelberg, Germany \\
$^{2}$Astronomy Department, Yale University, P.O.\ Box 208101, New Haven, CT 06520-8101, USA \\
$^{3}$Max Planck Institute for Astrophysics, Karl-Schwarzschild-Stra{\ss}e 1, 85748, Garching, Germany \\
}

\begin{document}

\date{Accepted 2018 October 05. Received 2018 September 28; in original form 2018 August 14}

\pagerange{\pageref{firstpage}--\pageref{lastpage}} \pubyear{2018}

\maketitle

\label{firstpage}

\begin{abstract}
\noindent The circumgalactic medium (CGM), i.e.\ the gaseous haloes around galaxies, is both the reservoir of gas that fuels galaxy growth and the repository of gas expelled by galactic winds. Most cosmological, hydrodynamical simulations focus their computational effort on the galaxies themselves and treat the CGM more coarsely, which means small-scale structure cannot be resolved. We get around this issue by running zoom-in simulations of a Milky Way-mass galaxy with standard mass refinement and \emph{additional uniform spatial refinement} within the virial radius. This results in a detailed view of its gaseous halo at unprecedented (1~kpc) uniform resolution with only a moderate increase in computational time. The improved spatial resolution does not impact the central galaxy or the average density of the CGM. However, it drastically changes the radial profile of the neutral hydrogen column density, which is enhanced at galactocentric radii larger than 40~kpc. The covering fraction of Lyman-Limit Systems within 150~kpc is almost doubled. We therefore conclude that some of the observational properties of the CGM are strongly resolution dependent. Increasing the resolution in the CGM, without increasing the resolution of the galaxies, is a promising and computationally efficient method to push the boundaries of state-of-the-art simulations. 

\end{abstract}

\begin{keywords}
methods: numerical -- hydrodynamics -- galaxies: evolution -- galaxies: formation -- galaxies: haloes -- intergalactic medium 
\end{keywords}

\section{Introduction} \label{sec:intro}
 
Cosmological, hydrodynamical simulations have seen dramatic advances in recent years, producing well-resolved stellar disks with sizes, scale lengths, and stellar masses in reasonable agreement with observations of late-type galaxies \citep[e.g.][]{Schaye2015, Grand2017, Garrison2017, Genel2018}. These simulations have wildly different feedback models, suggesting that galaxy properties may not discriminate between them. The circumgalactic medium (CGM; i.e.\ the gaseous haloes around galaxies) will enable us to study how galaxies' gas reservoirs are replenished and how galactic winds alter the galaxies and their environments. Studying this regime can also help test theoretical models by comparing them to available observations of the CGM \citep[e.g.][]{Putman2012, Tumlinson2017}. 

It is often assumed that the hydrodynamical processes outside the galaxy's interstellar medium (ISM) are well-resolved. However, the resolution in state-of-the-art simulations is generally adaptive in a (quasi-)Lagrangian sense, such that the mass resolution is kept fixed. The spatial resolution therefore drops quickly with galactocentric radius in the CGM, in lock-step with the much lower densities there. Worryingly, many idealized studies (which give up the cosmological context in favour of much higher spatial resolution) found that the properties of `halo gas' change considerably with improved resolution \citep[e.g.][]{Scannapieco2015, Schneider2017, Mandelker2018, McCourt2018, Sparre2018}. This suggests that the CGM in cosmological simulations is under-resolved, which could affect our ability to predict or reproduce CGM observables and also have major consequences for the amount of gas accretion on to galaxies and for the mixing of metals. 

Questions surrounding the physical and observable properties of the CGM, the role of feedback, the re-accretion of previously expelled gas, and the distribution of heavy elements would greatly benefit from simulations that offer higher spatial resolution in the CGM than permitted by standard simulation techniques. We therefore present zoom-in simulations centred on a Milky Way-mass galaxy taken from the `Auriga' project \citep{Grand2017}, resimulated with fixed spatial resolution within the gaseous haloes of all galaxies within the zoom-in region. A similar approach was taken by \citet{Miniati2014} to study turbulence in a merging galaxy cluster at fixed spatial resolution, achieving 10~kpc uniform spatial resolution in the intracluster medium.

In this letter, we present properties of the halo gas around a simulated Milky Way analogue with direct implications for observations of atomic hydrogen (also called ``neutral hydrogen'' or ``H\,\textsc{i}'') around the Milky Way. The new simulation refinement method is described in Section~\ref{sec:sim}. In Section~\ref{sec:results} we present our results on the density and H\,\textsc{i} structure in the CGM and we conclude in Section~\ref{sec:concl}.

\section{Method} \label{sec:sim}

This work is an extension of the Auriga project \citep{Grand2017}, which consists of a large number of zoom-in magnetohydrodynamical, cosmological simulations of isolated Milky Way-mass galaxies and their environments. These simulations produce realistic disc-dominated galaxies with stellar masses, galaxy sizes, rotation curves, star formation rates, and metallicities in agreement with observations. The simulations were carried out with the quasi-Lagrangian moving-mesh code \textsc{arepo} \citep{Springel2010}. This original suite employs a fixed mass resolution, which means that mesh cells within the zoom-in region are refined (i.e.\ split in two) or derefined (i.e.\ merged with their neighbours) if their mass differs by more than a factor of two from the target mass resolution. 

We resimulate, using \textsc{arepo}, one of the Auriga galaxies (halo 6) at standard mass resolution, i.e.\ the target cell mass for baryons is $5.4\times10^4$~M$_{\astrosun}$ and dark matter particle masses are $2.9\times10^5$~M$_{\astrosun}$. In two other simulations, we impose an additional refinement criterion based on the physical cell volume. In this case, a cell is also refined when its volume is more than twice the target volume resolution and only derefined if its mass \emph{and} volume are less than half the target resolution. In the ISM of the galaxy, the density is high, which means that the cells are smaller than the imposed spatial resolution. There is thus no change in the resolution of the galaxy itself, but the resolution is greatly improved in the lower density CGM. 

In order to select the region that is spatially refined, we run \textsc{subfind} \citep{Springel2001, Dolag2009} on-the-fly. Dark matter haloes are identified using a Friends-of-Friends (FoF) algorithm with a linking length $b=0.2$. Each spherical region of radius $1.2R_\mathrm{vir}$ centred on every FoF halo with $M_\mathrm{halo}>10^{8.7}$~M$_\odot$ is tagged for additional spatial refinement. In this work, we define the virial radius, $R_\mathrm{vir}$, as the radius within which the mean overdensity is 200 times the mean density of the Universe at its redshift. 

We create a \textsc{subfind} catalogue at each of the 128 output redshifts between $z=47$ and $z=0$, with time intervals of $4-204$~Myr, and `dye' cells within $1.2R_\mathrm{vir}$ of sufficiently massive haloes with a passive scalar of value unity. As the simulation continues between two consecutive outputs, the scalar dye is advected with the fluid and used as a marker for the CGM of the galaxies. At each computational time-step, a cell is spatially refined (as described in \citealt{Springel2010}) if its refinement scalar is larger than 0.9, ensuring that the virial regions remain well-resolved until the dye is reinitialized at the next output redshift. 

The three simulations presented here have
\begin{enumerate}
\item standard mass refinement, i.e.\ all cells contain approximately $5.4\times10^4$~M$_{\astrosun}$ of gas;
\item standard mass refinement plus fixed spatial refinement where $\rho<\rho_\mathrm{threshold}/512$, resulting in gas cell volumes of approximately 8~proper kpc$^3$ or spatial resolution $\approx2$~kpc;
\item standard mass refinement plus fixed spatial refinement where $\rho<\rho_\mathrm{threshold}/64$, resulting in gas cell volumes of approximately 1~proper kpc$^3$ or spatial resolution $\approx1$~kpc,
\end{enumerate}
where $\rho_\mathrm{threshold}$ is the gas density above which star formation occurs (i.e.\ $n_\mathrm{H}^\star=0.11$~cm$^{-3}$). 

Our simulations use the same physical model as the original Auriga suite (see \citealt{Grand2017}). Besides the additional spatial refinement, the only other significant difference with the original Auriga project is that we inject mass and metals ejected by a star particle only into its host cell, rather than into its 64 neighbours. However, the global galaxy properties are not affected by this choice. 

In this letter, we are mainly interested in properties of the CGM and we focus on the most common element: hydrogen. The neutral hydrogen fraction is calculated on-the-fly, using the ionization and recombination rates from \citet{Katz1996} and the ionizing ultra-violet (UV) background from \citet{Faucher2009}. Self-shielding from the UV background is implemented using the fits to the self-shielding fraction in radiative transfer simulations by \citet{Rahmati2013}. Gas above $\rho_\mathrm{threshold}$ is put on an artificial equation of state, because the simulations cannot resolve the multi-phase nature of the ISM. The density of the star-forming gas should therefore be treated with caution.

\section{Results}\label{sec:results}

\begin{table}
\begin{center}
\caption{\label{tab:prop} \small Properties of the galaxy and halo in our simulations at $z=0$: simulation refinement strategy, total stellar mass within 30~kpc from the centre ($M_\mathrm{star}$), total ISM mass within 30~kpc from the centre ($M_\mathrm{ISM}$), total CGM mass ($M_\mathrm{CGM}$), total H\,\textsc{i} mass in the CGM ($M_\mathrm{CGM}^{\rm H\,\textsc{i}}$), number of gas cells in the CGM ($N_\mathrm{CGM}^\mathrm{cell}$).}
\vspace{-4mm}
\begin{tabular}[t]{rrrrrr}
\hline \\[-3mm]
simulation & $M_\mathrm{star}$ & $M_\mathrm{ISM}$ & $M_\mathrm{CGM}$ & $M_\mathrm{CGM}^{\rm H\,\textsc{i}}$ & $N_\mathrm{CGM}^\mathrm{cell}$ \\
refinement & (M$_{\astrosun}$) &  (M$_{\astrosun}$) & (M$_{\astrosun}$) & (M$_{\astrosun}$) & \\
\hline \\[-4mm]                                                                                                                                       
mass only & $10^{10.73}$ & $10^{9.48}$ & $10^{10.97}$ & $10^{10.26}$ & 1.6M \\
+ 2~kpc    & $10^{10.72}$ & $10^{9.41}$ & $10^{10.91}$ & $10^{10.24}$ & 16.3M \\
+ 1~kpc    & $10^{10.67}$ & $10^{9.52}$ & $10^{10.96}$ & $10^{10.29}$ & 132.4M \\[-1mm]
\hline
\end{tabular}
\end{center}
\end{table}   
\begin{figure*}
\center
\includegraphics[scale=0.65]{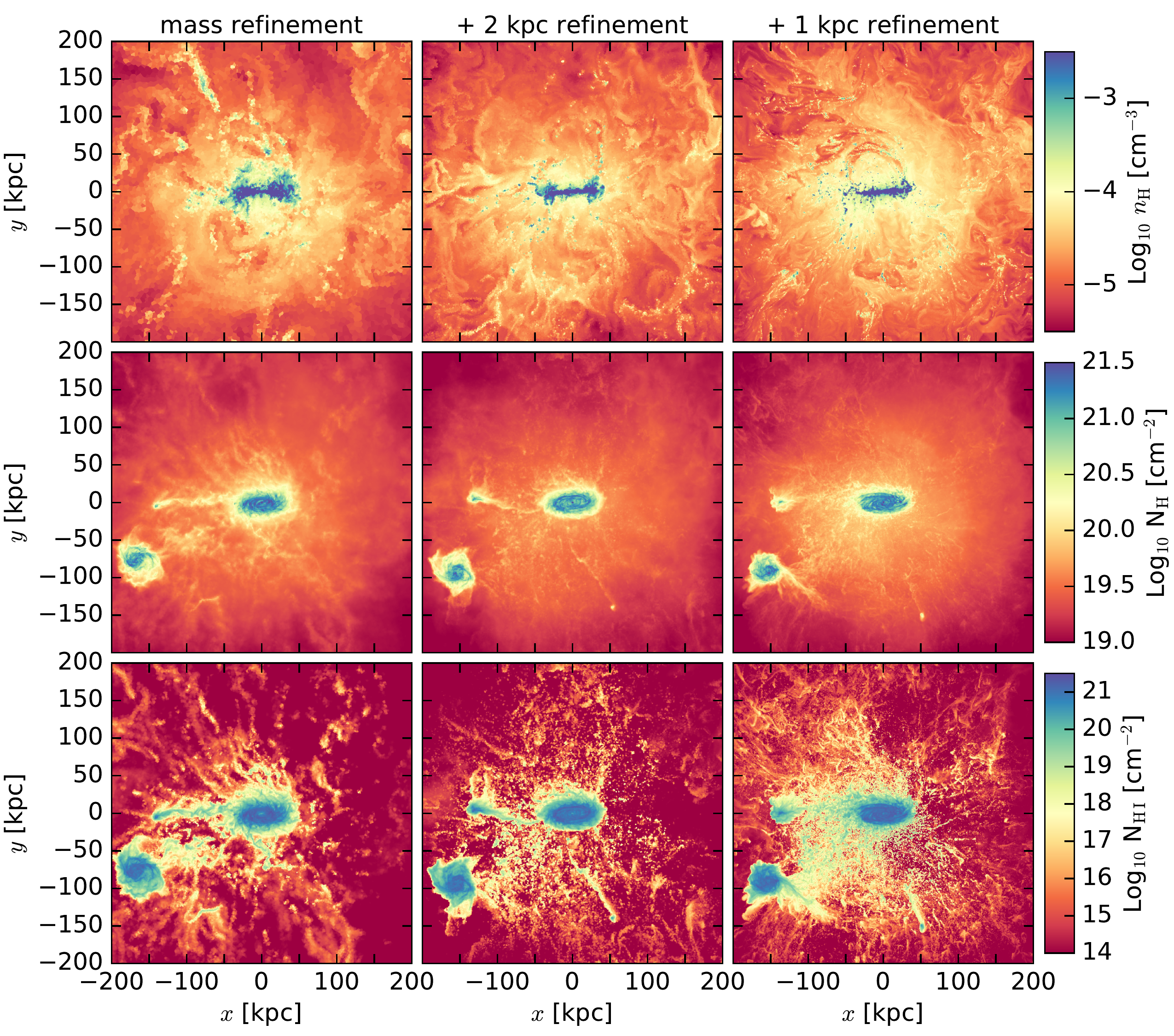}
\caption {\label{fig:img} $200\times200$~kpc$^2$ images of the gas in and around a Milky Way-mass galaxy at $z=0$ in a simulation with only mass refinement (left-hand panels), a simulation with additional spatial refinement of $\approx2$~kpc (middle panels), and a simulation with additional spatial refinement of $\approx1$~kpc (right-hand panels). Top panels: an infinitesimally thin slice of the hydrogen number density. Middle panels: total hydrogen column density in a 600~kpc column. Bottom panels: neutral hydrogen column density in a 600~kpc column. Higher resolution in the halo results in more small-scale structure, including more dense gas clumps and thin filaments. The covering fraction of high-column density H\,\textsc{i} is highest in the 1~kpc simulation.}
\vspace{-3mm}
\end{figure*} 
%

Gas accretes on to galaxies from the intergalactic medium after it has passed through the CGM. A better resolved CGM may exhibit different properties, which could affect cooling and thus the growth of the central galaxy. We find, however, that the global properties of this galaxy and its halo remain unchanged. The virial radius is approximately 337~kpc and varies by less than 0.5~per cent. Table~\ref{tab:prop} lists some of the global properties of the galaxy and its CGM, which is defined as all gas cells within $R_\mathrm{vir}$ with $\rho<\rho_\mathrm{thresh}$. The ISM of a galaxy is defined as all star-forming gas (i.e.\ $\rho>\rho_\mathrm{thresh}$) within 30~kpc of its centre. The CGM values include gas associated with satellites, while excluding their ISM, but this choice does not affect our conclusions. 

The stellar mass and ISM mass vary by only 0.06~dex and 0.11~dex, respectively, which is consistent with being due to chaotic behaviour \citep{Genel2019}. The bulge-to-disc ratio varies by just 1 per cent. The total mass in the CGM and the H\,\textsc{i} mass in the CGM also remain approximately the same when the resolution in the CGM is enhanced. However, these are all dominated by the dense gas in the central regions, where the spatial refinement is not in effect. H\,\textsc{i} makes up $\approx20$ per cent of the total gas mass in the CGM (and $\approx28$ per cent of the total hydrogen in the CGM). 

Compared to the simulation with only mass refinement, the number of resolution elements in the CGM in the 2~kpc and 1~kpc spatially refined simulations is enhanced substantially, by a factor of 10 and 82, respectively. The computational effort, however, only increases by a factor of 2~and~8, respectively. This new method is therefore cheap compared to increasing the resolution by using a higher mass resolution.
Simulations of this type are well suited for high-resolution studies of the gas in galaxy haloes and potentially in other regimes. When adding spatial refinement, most of the additional resolution elements are added to the low-density outer halo. The resolution increase within 150~kpc ($\approx0.45R_\mathrm{vir}$) is only a factor of 3 (19) for the 2~kpc (1~kpc) spatially refined simulations.

Fig.~\ref{fig:img} shows $200\times200$~kpc$^2$ images centred on the Milky Way-mass galaxy in our three simulations. The left-hand panels show a simulation with standard mass refinement ($5\times10^4$~M$_{\astrosun}$), which means that the resolution decreases with decreasing density. The middle and left-hand panels show simulations with the same mass refinement, but additional spatial refinement, resulting in a maximum cell volume of about (2~kpc)$^3$ and (1~kpc)$^3$, respectively.  

The top panels show the hydrogen number density in an infinitesimally thin slice through the centre of the halo. The CGM shows significantly more small-scale structure at high resolution, including a larger number of dense clumps and smaller turbulent eddies. The middle panels show the total hydrogen column density integrated over 600~kpc along the line-of-sight. Although the contrast in the CGM is not as large as for the density slice, because the fluctuations average out in projection, the amount of structure on small scales is still clearly enhanced in the spatially refined simulations. The neutral hydrogen column density is shown in the bottom panels. The small-scale structure of dense clumps and thin filaments is much more visible even though the dynamic range spans a much larger range. Also striking is that the average $N_\mathrm{H\,\textsc{i}}$ is higher in the 1~kpc spatial refinement simulation and that the covering fraction of high-column density systems is larger by a substantial amount. We quantify this result below. Some of the high-column density systems result from gas stripped from a satellite flyby, but most of the covering fraction is not associated with such events. 

\begin{figure}
\center
\includegraphics[scale=0.45]{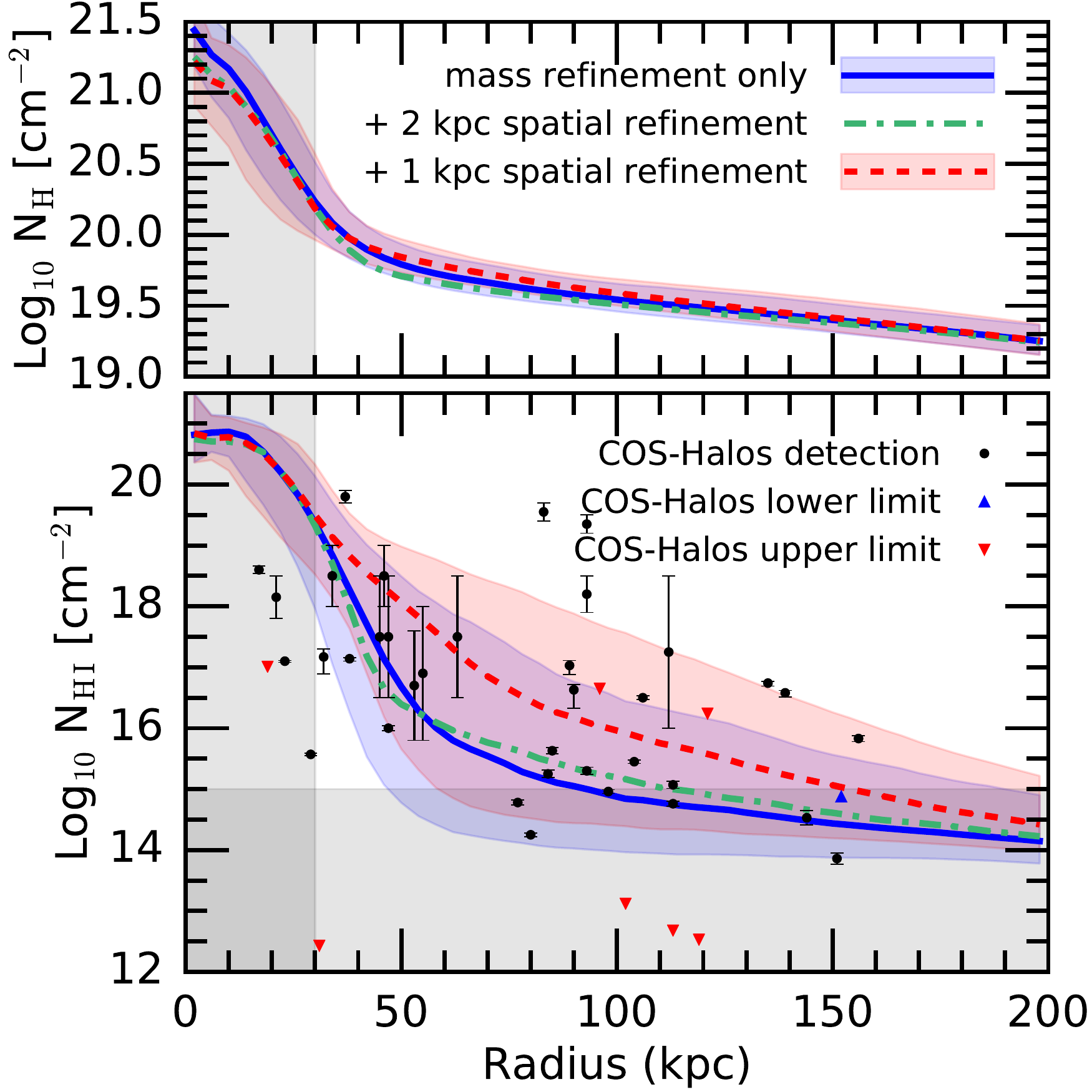}
\caption {\label{fig:rad} Median 2D radial profile for the total hydrogen column density (top panel) and neutral hydrogen column density (bottom panel) at $z=0.3-0$ in our three simulations, described in Section~\ref{sec:sim}. Shaded blue and red regions show the $1\sigma$ scatter for the mass refinement only simulation and for the 1~kpc spatially refined simulation. The bottom panel includes observations of H\,\textsc{i} absorption systems from the COS-Halos survey \citep{Tumlinson2013, Prochaska2017} at $z\approx0.2$, where black circles indicate detections, blue upward triangles indicate lower limits, and red downward triangles indicate upper limits. Regions that are shaded grey could be dominated by ISM gas or project multiple H\,\textsc{i} systems and should therefore be treated with caution. The total hydrogen column density (and thus also the density profile of the CGM) does not depend on the resolution of our simulation. However, the neutral hydrogen column density is much higher in the 1~kpc spatially refined simulation. The $N_\mathrm{H\,\textsc{i}}$ enhancement is largest between 40~and 150~kpc, by up to 1.6~dex.} 
\vspace{-3mm}
\end{figure} 

In Fig.~\ref{fig:rad}, we quantify the median projected two-dimensional surface density profile for all hydrogen (top panel) and for only neutral hydrogen (bottom panel) in a 600~kpc column. Note that the column density range in the bottom panel is much larger than in the top panel. The solid, blue curves are radial profiles for our standard mass refinement simulation, whereas the dot-dashed, green curves and dashed, red curves are based on simulations with additional spatial refinement of 2~kpc and 1~kpc, respectively. Shaded regions show the 16th and 84th percentiles of the distribution, or the $1\sigma$ scatter, for the lowest and highest resolution simulation. 
To limit the impact of satellites and stochasticity due to galactic winds, we show the median of 3 orthogonal projections in the 22 simulation outputs between $z=0.3$ and $z=0$, but this choice does not affect our conclusions. 

The central 30~kpc is dominated by the unresolved ISM and is shaded grey, because we cannot trust the column densities within this region. High-column density systems ($N_\mathrm{H\,\textsc{i}}\gtrsim10^{15}$~cm$^{-2}$) are rare, so a 600~kpc sightline will be dominated by a single neutral hydrogen system in the vast majority of cases. However, because H\,\textsc{i} systems with smaller column densities are so numerous, it becomes likely that a 600~kpc sightline intersects multiple clouds. For a fair comparison to observations in the low-column density regime, it is therefore necessary to create synthetic spectra and isolate individual H\,\textsc{i} systems, which we leave for future work. In Fig.~\ref{fig:rad}, $N_\mathrm{H\,\textsc{i}}<10^{15}$~cm$^{-2}$ is shaded grey to indicate that these column densities may be overestimated compared to observations. 

The total projected density profile is relatively shallow outside the galaxy, decreasing by one order of magnitude from 30 to 200~kpc. $N_\mathrm{H\,\textsc{i}}$, on the other hand, decreases by 5 orders of magnitude over the same radial range. This means that neutral hydrogen is much more centrally concentrated than the mass. Furthermore, there are only small differences in the median $N_\mathrm{H}$ between the 3 simulations and the scatter is small in all three simulations. This shows that the average CGM density is not affected by resolution effects. However, we calculated the average clumping factor, i.e.\ $\langle\rho^2\rangle / \langle\rho\rangle^2$, at $30-200$~kpc and find a 10~per cent (30~per cent) increase with additional 2~kpc (1~kpc) spatial refinement.

The H\,\textsc{i} column density does not change substantially with additional 2~kpc refinement, probably because the resolution in the central 150~kpc is only slightly higher than the standard simulation with only mass refinement. 
However, the median neutral hydrogen column density is much higher in the 1~kpc spatially refined simulation, by up to 1.6~dex. The H\,\textsc{i} column density could potentially increase even further, because the results are not yet converged. The scatter between H\,\textsc{i} sightlines is large in all simulations, especially compared to the scatter in the total hydrogen column, and does not appear to depend significantly on resolution. Interestingly, the 1~kpc spatially refined simulation reproduces the observed high-column density systems better than the standard simulation.

\begin{figure}
\center
\includegraphics[scale=0.45]{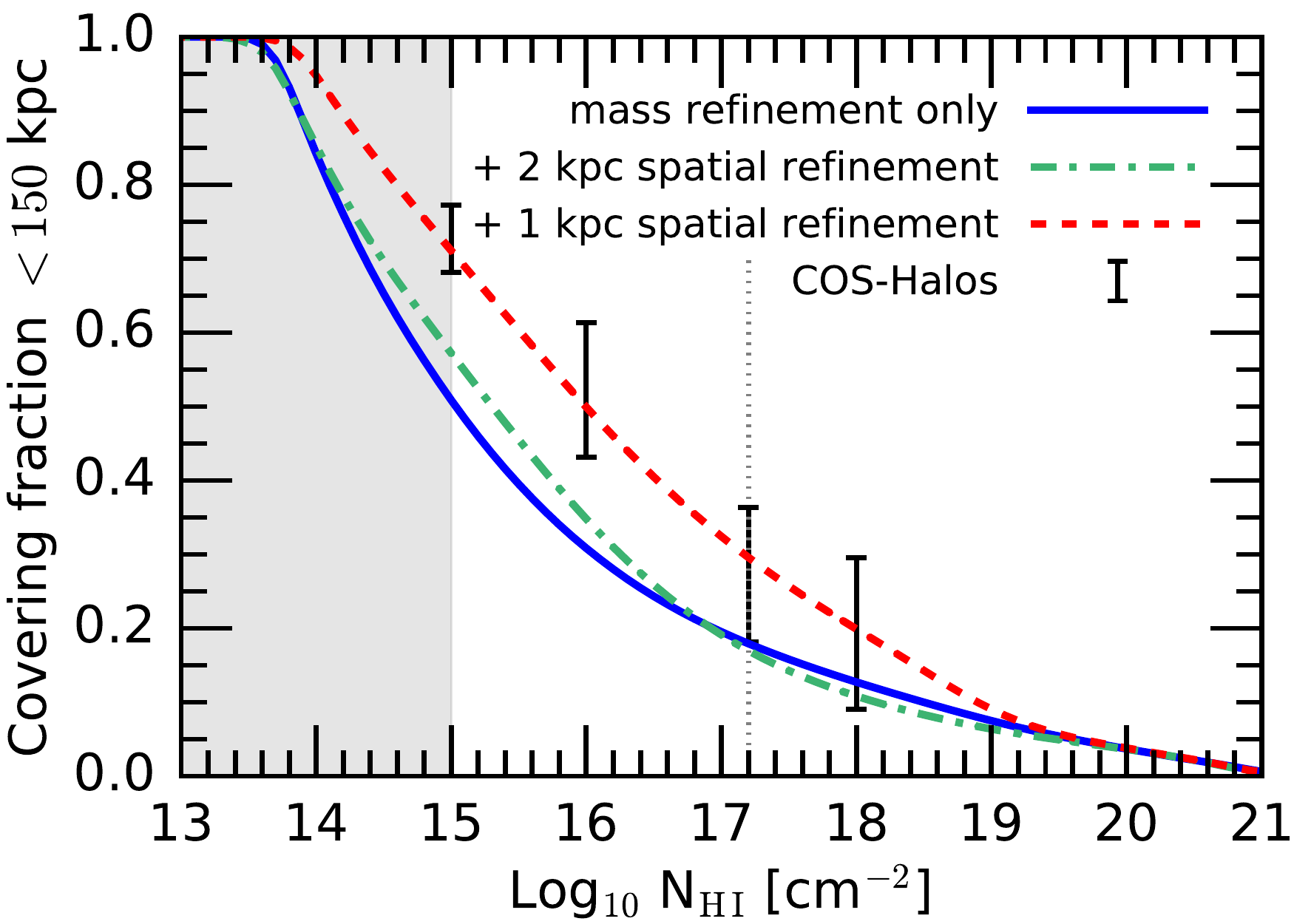}
\caption {\label{fig:fcov} The covering fraction of neutral hydrogen column density systems above the value indicated on the $x$-axis, within a projected radius of 150~kpc from the central galaxy, between $z=0.3$ and $z=0$. The line styles of the curves are identical to those used in Fig.~\ref{fig:rad}. The vertical dotted, grey line indicates the minimum column density for LLSs ($N_\mathrm{H\,\textsc{i}}\geq10^{17.2}$~cm$^{-2}$) and the black error bars show the covering fractions in the COS-Halos survey \citep{Tumlinson2013, Prochaska2017}. The covering fraction of all systems with $N_\mathrm{H\,\textsc{i}}\lesssim10^{19}$~cm$^{-2}$ is much higher with additional 1~kpc spatial refinement and the covering fraction of LLSs within 150~kpc increases from 18~to 30~per cent.}
\vspace{-3mm}
\end{figure} 

Another way to quantify the increase in high-column density systems with resolution is using covering fractions of gas with $N_\mathrm{H\,\textsc{i}}$ above a certain value. Fig.~\ref{fig:fcov} shows the covering fraction of H\,\textsc{i} column densities above the value indicated on the $x$-axis, within a projected radius of 150~kpc of the central galaxy. We use the same simulation outputs, projection axes, and line styles as in Fig.~\ref{fig:rad}. 

At the highest column densities, $N_\mathrm{H\,\textsc{i}}\gtrsim10^{20}$~cm$^{-2}$, the simulations exhibit very similar covering fractions. This is due to the fact that these column densities are dominated by dense gas for which the standard mass refinement already reduces the cell size to below 1~kpc, resulting in no difference between the simulations in this regime. However, at lower column densities, $N_\mathrm{H\,\textsc{i}}\lesssim10^{19}$~cm$^{-2}$, the additional spatial refinement becomes important and the covering fraction of these systems is much higher with 1~kpc spatial refinement. An increase in the H\,\textsc{i} covering fraction was also seen in simulations with increasingly better mass refinement, both in zoom-in simulations \citep{Faucher2016} as well as in lower resolution full cosmological volume simulations with a statistical sample of galaxies \citep{Rahmati2015}. 
This increase in the covering fraction does not mean the total amount of H\,\textsc{i} is increased substantially, as also demonstrated in Table~\ref{tab:prop}, because most of the H\,\textsc{i} mass is located in large H\,\textsc{i} discs around the galaxies and only roughly 1~per cent of the H\,\textsc{i} is located between 50~kpc and $R_\mathrm{vir}$ (excluding extended discs around satellites). 

Lyman-Limit Systems (LLSs) are those with $N_\mathrm{H\,\textsc{i}}>10^{17.2}$~cm$^{-2}$ (indicated by a vertical dotted, grey line). The COS-Halos observations, shown in Fig.~\ref{fig:rad}, exhibit a covering fraction of LLSs of $18-36$ per cent (including uncertainties due to $1\sigma$ errors and upper and lower limits), shown by a black error bar. The covering fraction of LLSs within 150~kpc rises from 18 per cent in the standard mass refined simulation to 30 per cent with additional 1~kpc spatial refinement. Overall, our highest spatial resolution simulation agrees best with the observations, especially at $N_\mathrm{H\,\textsc{i}}>10^{15-16}$~cm$^{-2}$. However, it remains to be seen what the covering fractions would be in a simulation with even higher resolution.

\section{Discussion and conclusions} \label{sec:concl}

We have presented a new refinement technique to simulate the CGM at uniform spatial resolution. This enables us to achieve much higher resolution at relatively low cost. We ran three cosmological simulations of the same Milky Way-mass halo: one with standard mass refinement, one with mass refinement and additional 2~kpc spatial refinement within $1.2R_\mathrm{vir}$, and one with additional 1~kpc spatial refinement. We find that the global properties of the galaxy are not affected by the better sampling of the halo gas. Similarly, the median density profile and its $1\sigma$ scatter are also robust to changes in the resolution of the halo gas. However, we find a large impact on the H\,\textsc{i} column density in the CGM. The median $N_\mathrm{H\,\textsc{i}}$ 2D radial profile is substantially higher (by up to 1.6~dex) in the simulation with additional 1~kpc spatial refinement as compared to the simulation with only mass refinement. As a result, the covering fraction of LLSs within 150 kpc from the galaxy centre increases from 18 to 30 per cent when including 1~kpc spatial refinement. The large reservoirs of cool CGM detected in observations \citep[e.g.][]{Werk2014} arise naturally in cosmological simulations.

However, there are some uncertainties remaining that affect the normalization of our results. The correction due to self-shielding as given in \citet{Rahmati2013} was derived from simulations with much lower resolution and may not be applicable to the small clouds we can resolve here. Therefore, we repeated our calculations in the optically thin regime. This reduces the column densities in $N_\mathrm{H\,\textsc{i}}\gtrsim10^{16}$~cm$^{-2}$ systems and strongly affects the normalization. However, qualitatively our results are unchanged and the simulations with 1~kpc spatial resolution have substantially higher $N_\mathrm{H\,\textsc{i}}$. The H\,\textsc{i} fraction would also be lower across the density range if our simulations used ionization and recombination rates given by \citet{Hui1997}, as done by \citet{Rahmati2013}, instead of following \citet{Katz1996}. This again affects the absolute normalization of our H\,\textsc{i} column density results, but none of the reported relative differences between our simulations. 

We conclude that the growth of the central galaxy and the bulk properties of the halo gas are not strongly affected by improved spatial resolution in the CGM. Simulations with only mass refinement therefore seem adequate to capture these aspects of galaxy evolution. However, observables that probe the extremes of the halo gas distribution, such as the low-temperature CGM, are strongly resolution dependent. Cosmological, hydrodynamical simulations with additional spatial refinement in the CGM can help bridge the gap between simulations with a realistic cosmological environment and high-resolution idealized simulations. Our results suggest that future observations with telescopes such as ASKAP, MeerKAT, and SKA are likely to detect more H\,\textsc{i} 21~cm emission than predicted by low-resolution cosmological simulations. In combination with observations, a well-resolved CGM will allow us to test, refine, and potentially rule out feedback models. In future work, we will study the impact of our improved uniform spatial resolution on metal-line emission and absorption in the CGM.

\section*{Acknowledgements}
This work is part of the HITS-Yale Program in Astrophysics (HYPA), sponsored by the Klaus Tschira Foundation. We would like to thank the referee for helpful comments as well as the Simons Foundation and the organizers and participants of the Simons Symposium `Galactic Superwinds: Beyond Phenomenology', for stimulating discussions and inspiration for this work. 
FvdB is supported by the National Aeronautics and Space Administration under Grant No. 17-ATP17-0028 issued through the Astrophysics Theory Program and by the US National Science Foundation through grant AST 1516962.

\bibliographystyle{mnras}
\bibliography{CGMrefine}

\bsp

\label{lastpage}

\end{document}